\documentclass[aps,twocolumn,prb]{revtex4}
\usepackage{graphicx}

\begin{document}
% You should use BibTeX and apsrev.bst for references
%\bibliographystyle{apsrev}

%\preprint{}

\title{Algebraic derivation of spectrum of the Dirac Hamiltonian for arbitrary
 combination of Lorentz-scalar and Lorentz-vector Coulomb potentials }

% Optional argument for running titles on pages
%\title[]{}


% other informatio
\author{Tamar~T.~Khachidze and Anzor~A.~Khelashvili}

%\thanks{}
\affiliation{Department of Theoretical Physics and IHEPI of Ivane
Javakhishvili Tbilisi State University,
                         I.Chavchavadze ave. 3, 0128, Tbilisi, Georgia}
%\email{khelash@ictsu.tsu.edu.ge}


\date{\today}

\begin{abstract}

                   Spectrum of the Dirac Equation is obtained algebraically for arbitrary combination of
                  Lorentz-scalar and Lorentz-vector Coulomb potentials using the Witten's Superalgebra
                  approach. The result coincides with that, known from the explicit solution of Dirac
                  equation.

\end{abstract}

%\date{\today}

\maketitle

 Recently \cite{Khachidze2006} we have derived the dynamical symmetry operator
    for the Dirac Hamiltonian in case of arbitrary combination of  Lorentz-scalar
     and Lorentz-vector Coulomb potentials. The operator of this kind was known
     earlier \cite{Leviatan2004} , but our derivation seems to be very simple and
     more transparent. Our consideration concerns to general central
     potentials $V(r)$
     and $S(r)$  in the Dirac equation  and under requirement the anticommutativity of looked
      operator with the Dirac operator

      \begin{equation}
      K=\beta\left( \vec{\Sigma} \cdot \vec{l} +1 \right)
      \end{equation}
and simultaneous commutativity with the Dirac Hamiltonian

 \begin{equation}
     H=\vec{\alpha} \cdot \vec{p}+\beta m +V(r)+ \beta S(r)
 \end{equation}
we have obtained the following expression for this operator
 \begin{equation}
      X=\gamma^5 (\vec{\alpha} \cdot
        \hat{\vec{r}})(ma_V+Ha_S) + \imath \gamma^5 K
      (H-\beta m)
      \end{equation}
Here   $a_V,_S$ are strengths of corresponding Coulomb potentials.

      It is worthwhile to underline, that the commutativity (i.e. symmetry) with the Dirac Hamiltonian (2)
      follows only both for Coulomb potentials, as a result  of degeneracy with respect of two signs of eigenvalues
      of $K$-operator, $\pm\kappa$ .

     In present paper we want to obtain spectrum of the
      Dirac Hamiltonian pure algebraically, without any referring
      on equations of motion. Our method is based on Witten's superalgebra, which follows immediately
      as soon as an anticommuting with $K$ operator is constructed. It is only sufficient to identify SUSY
       generators
       as follows:
        \begin{equation}
      Q_1=X, ~~~ Q_2=\imath \frac{XK}{|\kappa|}
      \end{equation}

   Then it follows from anticommutativity $\{X,K\}=0$ , that

    \begin{equation}
      \{Q_1,Q_2\}=0, ~~~~ Q_1^2=Q_2^2\equiv h
      \end{equation}

Therefore we are faced to Witten's algebra $S(2)$ , where $h=X^2$
is a Witten's Hamiltonian.

    Now we explore this algebra in a manner as in paper \cite{Katsura2006}.  One defines a SUSY ground state $|0\rangle $:
          \begin{equation}
     h|0\rangle = X^2 |0\rangle =0~~\longrightarrow~~X|0\rangle =0
      \end{equation}
    Because  $X^2$ is a square of Hermitian operator, it has a positive definite spectrum and one
    is competent to take zero this operator itself in ground state. By this requirement we'll
     obtain Hamiltonian in this ground state and, correspondingly, ground state energy. After that by well
      known ladder procedure we can construct the energies of all excited levels. We believe that this method
      requires more strong justification, but nevertheless we are convinced, that it is true.

     Let equate $X=0$ and solve from Eq. (3):

      \begin{equation}
      H=m [(\vec{\alpha}\cdot \hat{\vec{r}})a_S + \imath
      K]^{-1} [\imath K \beta - a_V (\vec{\alpha}\cdot
      \hat{\vec{r}})]=\frac{m}{\kappa^2+a_S^2} N
      \end{equation}
where

 $$
    N \equiv [(\vec{\alpha}\cdot\hat{\vec{r}})a_S-\imath K]
    [\imath K\beta - a_V (\vec{\alpha}\cdot \hat{\vec{r}})]=
            $$

\begin{equation}
= -a_S a_V + K[ K\beta+\imath a_V
     ( \vec{\alpha} \cdot \hat{\vec{r}})] - \imath a_S K \beta
     (\vec{\alpha}\cdot\hat{\vec{r}})
     \end{equation}

 Now we try to diagonalise this operator using
Foldy-Wouthuysen ~\cite{Foldy1950} like transformation. Because
the second and third terms do not commute with each others we need
several (at least two) such transformations.

   We choose the first transformation in the following manner

    \begin{equation}
      \exp(\imath S_1)= \exp \left( -\frac{1}{2} \beta
      (\vec{\alpha}\cdot \hat{\vec{r}}) w_1 \right)
      \end{equation}

It is evident that

 \begin{equation}
\exp(\imath S_1) (\vec{\alpha}\cdot \hat{\vec{r}}) \exp(-\imath
S_1)=\exp(2\imath S_1)(\vec{\alpha}\cdot \hat{\vec{r}})
\end{equation}

$$ \exp(\imath S_1) \beta \exp(-\imath S_1) = \exp(2 \imath S_1)
\beta $$

Moreover

$$\exp(\imath S_1) K \exp(-\imath S_1)=K,$$
\begin{equation}
 \exp(\imath S_1) \beta
K \exp(-\imath S_1) = \exp(2 \imath S_1) \beta K
\end{equation}

and

    \begin{equation}
\exp(\imath S_1) \beta (\vec{\alpha}\cdot \hat{\vec{r}})
\exp(-\imath S_1)=\beta (\vec{\alpha}\cdot \hat{\vec{r}})
      \end{equation}

   Therefore the first transformation acts as

$$N' \equiv \exp(\imath S_1) N
 \exp(- \imath S_1) =$$
 \begin{equation}
 = - a_S a_V + K \exp(2\imath S_1 )
[K \beta + \imath a_V(\vec{\alpha} \cdot \hat{\vec{r}} )]
 -\imath a_S K\beta(\vec{\alpha}\cdot\hat{\vec{r}})
     \end{equation}

    But  $\exp( 2\imath S_1)=
     \cosh {w_1}
     +\imath\beta(\vec{\alpha}\cdot\hat{\vec{r}})\sinh{w_1}$.
    Make use of this relation, we have

$$\exp(2\imath S_1) [K \beta +\imath a_V ( \vec{\alpha}\cdot
\hat{\vec{r}})]=$$
\begin{equation} \beta[K\cosh w_1 + a_V \sinh w_1]+ K
(\vec{\alpha}\cdot \hat {\vec{r}})[\imath a_V \cosh w_1 +\imath K
\sinh w_1 ]
\end{equation}
   Now in order to get rid of non-diagonal  $(\vec{\alpha}\cdot \hat {\vec{r}})$ terms, we must choose

\begin{equation}
\tanh w_1 = - \frac {a_V}{K}
\end{equation}

  Using simple trigonometric relations we arrive at
\begin{equation}
\exp(2\imath S_1)[K+\imath a_V (\vec{\alpha}\cdot \hat {\vec{r}})]
= K^{-1} \beta \sqrt {\kappa^2 - a_V^2}
\end{equation}

Let us perform the second F.-W. transformation

\begin{equation}
N''=\exp(\imath S_2) N' \exp(-\imath S_2),  ~~~~ where ~~~~ S_2 =
-\frac{1}{2}(\vec{\alpha}\cdot \hat {\vec{r}})w_2
\end{equation}

Now
 \begin{equation}
\exp(\imath S_2) K \beta \exp(-\imath S_2) = \exp(2\imath S_2) K
\beta
\end{equation}

$$\exp(\imath S_2) K \beta (\vec{\alpha}\cdot \hat
{\vec{r}})\exp(-\imath S_2) = \exp(2\imath S_2) K \beta
(\vec{\alpha}\cdot \hat {\vec{r}}) $$

$$ \exp(2\imath S_2) = \cos w_2 - \imath (\vec{\alpha}\cdot \hat
{\vec{r}}) \sin w_2$$
 Therefore

$$N'' = - a_S a_V +K \sqrt{\kappa^2 -
 a_V^2}\exp(2 \imath S_2 )\beta - $$
$$ - \imath a_S \exp(2\imath S_2) K \beta(\vec{\alpha}\cdot \hat
{\vec{r}})$$

$$ = -a_S a_V +K \sqrt{\kappa^2 - a_V^2}\beta \cos w_2 +\imath K
\sqrt{\kappa^2 - a_V^2} \beta(\vec{\alpha}\cdot \hat {\vec{r}})
\sin w_2 -$$
\begin{equation}
 -\imath a_S K \beta (\vec{\alpha}\cdot \hat {\vec{r}})
\cos w_2 + a_S K \beta \sin w_2
\end{equation}

Requiring absence of  $(\vec{\alpha}\cdot \hat {\vec{r}})$ terms
we have

 \begin{equation}
\tan w_2 = \frac {a_S}{\sqrt{\kappa^2-a_V^2}}
\end{equation}

Therefore

 \begin{equation}
N'' = - a_S a_V + K \beta \sqrt{\kappa^2-a_V^2+a_S^2}
\end{equation}

and finally

 \begin{equation}
H = \frac{m}{\kappa^2+a_S^2}\{ -a_S a_V + K \beta
\sqrt{\kappa^2-a_V^2+a_S^2}\}
\end{equation}

For eigenvalues in ground state we have

 \begin{equation}
E_0 = \frac{m}{\kappa^2+a_S^2}\left[ -a_S a_V  \pm \kappa
\sqrt{\kappa^2-a_V^2+a_S^2}\right]
\end{equation}

Now let us remember the result obtained by explicit solution of
the Dirac equation for this case \cite{Greiner1985}

$$\frac{E}{m}= \frac{-a_S a_V}{a_V^2+(n-|\kappa|+\gamma)^2}$$
 \begin{equation}\pm
\sqrt{\left(\frac{a_S a_V}{a_V^2+(n-|\kappa|+\gamma)^2}\right)^2+
\frac{(n-|\kappa|+\gamma)^2-a_S^2}{a_V^2+(n-|\kappa|+\gamma)^2}},
\end{equation}

where
\begin{equation}
\gamma^2=\kappa^2-a_V^2+a_S^2
\end{equation}

     In the ground state  $n=1$, $j=1/2$ $\longrightarrow$ $|\kappa|=j+1/2=1$, there remains

     \begin{equation}
E_0=m\left[\frac{-a_S a_V}{a_V^2+\gamma^2} \pm
\sqrt{\left(\frac{a_S a_V}{a_V^2+\gamma^2}\right)^2 + \frac
{\gamma^2 - a_S^2}{a_V^2+\gamma^2}}\right],
\end{equation}

which after obvious manipulations reduces to our above derived expression (23).
     Therefore by only algebraic methods we have obtained the correct expression for ground state energy.
     For obtaining of total spectrum it is sufficient now to use the Witten's algebra. Following the paper
      \cite{Katsura2006} ,
the ordinary ladder procedure consists in change (for our case):

$$ \gamma \longrightarrow \gamma + n - |\kappa| $$

    Making this, it follows from our lowest energy formula (23) the correct expression for total energy spectrum, eq. (24).

         In conclusion, by using of pure algebraic manipulations we obtained the spectrum of generalized Coulomb
    problem of the Dirac equation for arbitrary combination of Lorentz-scalar and Lorentz-vector potentials.
    This fact demonstrates a power of symmetry considerations, while method developed here requires more strong
     justification, which will be made in future investigations.

         The Authors are indebted  to express their gratitude to Prof. A.N.Tavkhelidze for many valuable discussions
         and critical remarks. We thank also Prof. Laszlo Jenkovsky for his  permanent attention in the period and
         after the Yalta Conference.

     This work was supported by NATO Reintegration Grant No. FEL.
     REG. 980767.

\end{document}